\documentclass{elsart}

\usepackage{amssymb}
\usepackage{amsmath}

\newcommand{\qln}{ \ln_{q} }
\newcommand{\qexp}{ \exp_{q} }
\newcommand{\ave}[1]{\left\langle #1 \right\rangle}

\begin{document}
\begin{frontmatter}

\title{Connections between Tsallis' formalisms employing
the standard linear average energy
and ones employing the normalized $q$-average energy}
\author[wada,polito]{T. Wada} and %\corauthref{cor1}}
\ead{wada@ee.ibaraki.ac.jp}
\author[polito,INFM]{A.M. Scarfone}
\ead{antonio.scarfone@polito.it}
\address[wada]{Department of Electrical and Electronic Engineering, 
Ibaraki University, Hitachi,~Ibaraki, 316-8511, Japan}
\address[polito]{Dipartimento di Fisica,
Politecnico di Torino, Corso Duca degli Abruzzi 24, 10129 Torino, Italy}
\address[INFM]{Istituto Nazionale di Fisica della Materia (INFM)\\
Politecnico di Torino, Corso Duca degli Abruzzi 24, 10129 Torino, Italy}

%\corauth[cor1]{Corresponding author.}

\begin{abstract}
Tsallis' thermostatistics with the standard linear average energy
is revisited by employing $S_{2-q}$, which is the Tsallis entropy with $q$ 
replaced by $2-q$. We explore the connections 
among the $S_{2-q}$ approach and the other different versions
of Tsallis formalisms.
It is shown that the normalized $q$-average energy
and the standard linear average energy are related to each other.
The relations among the Lagrange multipliers of the different versions
are revealed.
The relevant Legendre transform structures concerning the Lagrange
multipliers associated with the normalization of probability 
are studied.
It is shown that the generalized Massieu potential 
associated with $S_{2-q}$ and the linear average energy
is related to one associated with
the normalized Tsallis entropy and the normalized $q$-average energy.
\end{abstract}
\end{frontmatter}

%\thanks{current affilliation: Departimento di Fisica, 
%Politecnico di Torino, 10129 Torino, Italy}}

\section{Introduction}
\vspace*{-7mm}

Nowadays Tsallis' thermostatistics \cite{Tsallis88,Curado91,Tsallis98} is 
considered as one of the generalizations of the standard 
thermostatistics \cite{Callen} based on the Tsallis entropy
$S_q \equiv \sum_i(p_i-p_i^q) / (q-1)$, where $p_i$ stands for
a probability of $i$-th state and $q$ is a real parameter.
For the sake of simplicity the Boltzmann constant is set to unity 
throughout this paper. 
In the $q \to 1$ limit, $S_q$ reduces to the standard Boltzmann-Gibbs
(BG) entropy $S = -\sum_i p_i \ln (p_i)$.\\
During the last decade,
there have been vast numbers of basic studies and 
applications \cite{NEXT01,NEXT03,Refs},
and the formalism of Tsallis' thermostatistics has been evolved.
Tsallis' entropy was originally introduced \cite{Tsallis88}
with the standard average energy $U^{\rm (1)} = \sum_i p_i E_i$ 
as the internal energy constraint in the MaxEnt procedure.
(Here and hereafter, we use the superscript ${\rm (}i {\rm )}$ with 
$i=1,2,3$ in order to distinguish the three different 
average energies in Tsallis' thermostatistics.)\\
The second version \cite{Curado91} was proposed by replacing 
the energy constraint $U^{\rm (1)}$
with the unnormalized $q$-average energy $U_q^{\rm (2)} = \sum_i p_i^q E_i$
in order to restore the thermodynamic stability for all values of $q$
at the expense of the invariance of the probability distribution 
function (pdf) under the uniform translation of energy spectrum.
\\
The role of energy constraints ($U^{\rm (1)}, U_q^{\rm (2)},$ and 
$U_q^{\rm (3)}$) within Tsallis' thermostatistics
was precisely studied \cite{Tsallis98}, and the third (current) 
version was proposed by replacing the definition of the
energy constraint with the normalized 
$q$-average energy $U_q^{\rm (3)} = \sum_i p_i^q E_i / \sum_j p_j^q$,
which is also expressed as the average energy w.r.t. the
{\it so-called} escort probability $P_i \equiv p_i^q / \sum_j
p_j^q$ \cite{escort}.
Consequently Tsallis' thermostatistics has the two types of probabilities 
($p_i$ and $P_i$), 
which coincide with each other in standard thermostatistics ($q=1$).
\\
The correspondence between the two types of probabilities leads to
the so called ``$q \leftrightarrow 1/q$''-duality \cite{Tsallis98,Naudts-CSF}.
Raggio \cite{Raggio} had already shown that the equivalence 
between the first and third versions of Tsallis' formalism by utilizing 
the ``$q \leftrightarrow 1/q$''-duality, i.e., maximizing 
$S_q$ under the energy constraint of $U_q^{\rm (3)}$ is equivalent
to maximizing $S_{1/q}$ under that of $U^{\rm (1)}$.
\\
Through the efforts \cite{0th-law,Abe-Martinez01} to generalize the zeroth law 
of thermodynamics
within Tsallis' thermostatistics, it was revealed that the inverse
temperature is not simply the Lagrange multiplier associated with
the energy constraint.
For this reason, the Tsallis variational problem and the Legendre transform
structures have been extensively
studied by e.g., so-called {\it optimal Lagrange multiplier} (OLM)
formalism \cite{Martinez00,Abe01,Casas02}.
\\
In the literature, some derivatives of Tsallis' entropy have been proposed. 
One of them is the normalized Tsallis entropy 
$S_q^{\rm N} \equiv S_q / \sum_j p_j^q$ \cite{Landsberg,Rajagopal}, 
and another is 
the escort Tsallis entropy $S_q^{\rm E}$ \cite{escort-Sq}, which
emerges from $S_q$ by expressing $p_i$ in terms of the escort
probability $P_i$ and then renaming $P_i$ to $p_i$.
\\
Since Tsallis' thermostatistics has been still under development,
there remain some fundamental questions to be answered.
One of them is the choice of the energy average
which is used in the MaxEnt procedure of Tsallis' thermostatistics. 
Until now there are two main different opinions: one is to employ
the standard average energy $U^{\rm (1)}$; the other is
to employ the normalized $q$-average energy $U_q^{\rm (3)}$.
\\
Abe and Bagci \cite{Abe04} have shown that the generalized
relative entropy associated with $U_q^{\rm (3)}$ has nice properties,
which are superior to those associated with $U^{\rm (1)}$.
Di Sisto {\it et al.} \cite{Sisto} and Bashkirov \cite{Bashkirov}
have independently shown that the modified treatment of the
variational problem for the first version of Tsallis' thermostatistics
leads to the pdf which is analogous to the pdf of the third version.
\\
There exists another duality which is 
called ``$q \leftrightarrow 2-q$''-duality \cite{Naudts02}
in the $q$-deformed functions.
Baldovin and Robledo \cite{Baldovin} have observed that
the maximization of $S_{2-q}$ with the standard constraints
$\sum_i p_i E_i = U^{\rm (1)}$ and $\sum_i p_i = 1$ leads
to the $q$-exponential pdf. They suggested,
based on  the ``$q \leftrightarrow 2-q$''-duality, that the
mutual $S_q$ and $S_{2-q}$ elegantly generalize the standard
BG entropy, and pointed out that
some features are equally expressed by both $S_q$ and $S_{2-q}$,
but some others appears only via the use of either $S_q$ or $S_{2-q}$.
Finally, in Ref. \cite{Naudts@NEXT03} Naudts has analyzed both 
dualities of Tsallis' 
thermostatistics based on his generalized thermostatistics 
\cite{Naudts02,Naudts03}, and proposed
to replace $S_q$ with $S_{2-q}$ instead of introducing the normalized
$q$-average energy $U_q^{\rm (3)}$.

The purpose of this paper is twofold. Firstly, in order to study Naudts'
proposition, and to understand
a deeper relation between the formalisms employing $U^{\rm (1)}$ and 
the formalisms employing $U_q^{\rm (3)}$, we revisit 
Tsallis' thermostatistics
with $U^{\rm (1)}$ by using $S_{2-q}$ instead of $S_q$. 
The relationships among the 
Lagrange multipliers for the different
versions are obtained.\\
Secondly, we study a generalization of Massieu's potential associated with 
either $S_{2-q}$ and $U^{\rm (1)}$ or
$S_q^N$ and $U_q^{\rm (3)}$. 
The basic thermodynamic relations for these generalized 
potentials are discussed.\\
The plane of the paper is the following. In the next section we begin with the pdf 
obeying Tsallis' $q$-exponential, which maximizes $S_{2-q}$ under 
the constraint of the linear energy $U^{\rm (1)}$.
It is shown that the original pdf is equivalent to that of the 
(modified) first version with $q$ replaced by $2-q$.
In section 3, the escort pdf is introduced.
By utilizing the averages w.r.t. the escort pdf,
the original pdf is also shown to be equivalent to that of 
the third version and to that of the version which uses 
the normalized Tsallis entropy \cite{Landsberg,Rajagopal}. 
It is found 
that the standard linear average energy $U^{\rm (1)}$ and 
the normalized q-average 
energy $U_q^{\rm (3)}$ are related to each other as well as the corresponding 
Lagrange multipliers. In section 4, following the method used by Naudts 
to obtain the generalized free-energy \cite{Naudts@NEXT03,Naudts03},
we obtain a generalization of the Massieu's potential \cite{Callen} 
associated to 
the different formalisms. The final section is devoted to our conclusion.

%%%%%%%%%%%%%%%%%%%%%%%%%%%%%%%%%%%%%%%%%%%%%%%%%%%%%%%%%%%%%%%%%%%%%%%%%%%
\section{$S_{2-q}$ approach}
\vspace*{-7mm}

Following the route developed by  Naudts \cite{Naudts03} and Abe \cite{Abe}
independently, 
we can obtain the generalized entropy optimized by 
a given pdf.
We here take a similar approach \cite{Kaniadakis,Kaniadakis04} in order to 
obtain the generalized entropy
which is maximized by the Tsallis $q$-exponential pdf under
the constraint of the linear average energy $U^{\rm (1)}$.
Let us begin with the following $q$-exponential pdf
\begin{equation}
  p_i = \alpha \qexp \left( -\beta^{\rm (1)} E_i - \gamma^{\rm (1)} \right),
 \label{pdf}
\end{equation}
where $\alpha, \beta^{\rm (1)}$ and $\gamma^{\rm (1)}$ are real parameters 
to be determined later.
The Tsallis $q$-deformed exponential function $\qexp(x)$ 
\cite{Tsallis88,NEXT01,NEXT03} is defined by
\begin{equation}
  \qexp(x) \equiv \left( 1+(1-q)x \right)^{\frac{1}{1-q}},
  \label{qexp}
\end{equation}
where $q$ is a real parameter which characterizes the deformation.
The inverse function of $\qexp(x)$ is the $q$-logarithmic function
defined by
\begin{equation}
  \qln(x) \equiv \frac{x^{1-q}-1}{1-q}.
  \label{qln}
\end{equation}
We choose the parameter $\alpha$ so that
\begin{align}
  \frac{d}{d x} \big\{ x  \qln(x) \big\}
 =\qln \left( \frac{x}{\alpha} \right).
  \label{sc-log}
\end{align}
Then the parameter $q$ is related with
$\alpha$ by
\begin{equation}
  \frac{1}{\alpha} = (2-q)^{\frac{1}{1-q}}.
   \label{alpha}
\end{equation}
In addition, from Eq. \eqref{pdf}, we readily see that
\begin{equation}
  \qln \left( \frac{p_i}{\alpha} \right) 
        = -\beta^{\rm (1)} E_i - \gamma^{\rm (1)}.
  \label{scaled}
\end{equation}
This relation and the property of Eq. \eqref{sc-log} guarantee that
the pdf given in Eq. \eqref{pdf} is the solution of the following 
MaxEnt procedure \cite{Naudts03,Abe,Kaniadakis,Kaniadakis04}
\begin{equation} 
 \frac{\delta}{\delta p_i} 
 \left(
    S_{2-q} - \beta^{\rm (1)} \sum_j p_j E_j - \gamma^{\rm (1)} \sum_j p_j
 \right) = 0,
  \label{MaxEnt}
\end{equation}
where
\begin{equation} 
 S_{2-q}= \frac{\sum_i \left( p_i^{2-q} -p_i \right)}{q-1} =
-\sum_i p_i \qln (p_i),
  \label{S_2-q}
\end{equation}
is the Tsallis entropy with $q$ replaced by $2-q$.
We now see that the parameter $\beta^{\rm (1)}$ is the Lagrange 
multiplier associated with the linear average energy
\begin{equation}
  U^{\rm (1)} \equiv \sum_i p_i E_i,
\end{equation}
and $\gamma^{\rm (1)}$ is that associated with the normalization of the pdf,
$\sum_i p_i = 1$.

From Eqs. \eqref{sc-log}, \eqref{scaled} and \eqref{S_2-q}, we find
\begin{align}
  \frac{d S_{2-q}}{d \beta^{\rm (1)}} &= 
 -\sum_i \frac{d}{d p_i}\left( p_i \qln (p_i) \right)
\frac{d p_i}{d \beta^{\rm (1)}} 
  = - \sum_i \qln \left( \frac{p_i}{\alpha} \right) 
  \frac{d p_i}{d \beta^{\rm (1)}} \nonumber \\
  &=  \sum_i \left( \beta^{\rm (1)} E_i + \gamma^{\rm (1)} \right) 
         \frac{d p_i}{d \beta^{\rm (1)}}
 = \beta^{\rm (1)} \frac{d U^{\rm (1)}}{d \beta^{\rm (1)}},
  \label{S2-q-U1}
\end{align}
under the 'no work' condition, i.e., $d E_i = 0, \forall E_i$.
In the last step we used $\sum_i (d p_i / d\beta^{\rm (1)}) = 0$, which 
follows from the normalization of $p_i$.
We then obtain the thermodynamic Legendre relation \cite{Plastino97,Yamano}
\begin{equation}
  \frac{d S_{2-q}}{d U^{\rm (1)}} = \beta^{\rm (1)}.
\end{equation}

At this point, let us confirm Eq. \eqref{pdf} is equivalent to 
the pdf of the first version \cite{Tsallis88,Tsallis98,Sisto,Bashkirov} in 
Tsallis' thermostatics.
By taking the average of Eq. \eqref{scaled} w.r.t. $p_i$, we have
\begin{equation}
  \ave{ \qln \left( \frac{p_i}{\alpha}  \right) }
  = -\beta^{\rm (1)} U^{\rm (1)} -\gamma^{\rm (1)},
  \label{ave}
\end{equation}
and the l.h.s. is further expressed as
\begin{equation}
  \ave{ \qln \left( \frac{p_i}{\alpha}  \right) }
  = 1-(2-q) S_{2-q}.
\end{equation}
By combining the above two equations, we obtain
\begin{equation}
  \gamma^{\rm (1)} = (2-q) S_{2-q} -1 -\beta^{\rm (1)} U^{\rm (1)}.
  \label{gamma}
\end{equation}
Substituting this $\gamma^{\rm (1)}$ into  Eq. \eqref{pdf} and utilizing
the identity
\begin{equation}
   \qexp(x + y) = \qexp(x) \cdot \qexp\left( \frac{y}{1+(1-q)x} \right),
   \label{id-eq}
\end{equation}
it follows
\begin{align}
  p_i &=
 \alpha \qexp \big( 1-(2-q) S_{2-q} \big) \cdot
   \qexp \left( \frac{-\beta^{\rm (1)} (E_i-U^{\rm (1)}) }
   {1+(1-q) [ 1-(2-q) S_{2-q}] } \right)
  \label{pdf1} 
\nonumber \\
&= \frac{1}{\bar{Z}_{2-q}^{\rm (1)}} \cdot
   \qexp\left( \frac{-\beta^{\rm (1)} (E_i - U^{\rm (1)})}
      {(2-q)\sum_i p_i^{2-q}} \right),
\end{align}
where we introduced the generalized partition function as
\begin{equation}
  \bar{Z}_{2-q}^{\rm (1)} \equiv
 \left[ \alpha \qexp \left( 1-(2-q) S_{2-q} \right) \right]^{-1} =
 \left( \sum_i p_i^{2-q} \right)^{\frac{1}{q-1}}.
 \label{Z1}
\end{equation}
From the normalization of $p_i$, it follows that
\begin{equation}
\bar{Z}_{2-q}^{\rm (1)} =
\sum_i \qexp \left( \frac{-\beta^{\rm (1)} (E_i - U^{\rm (1)})}
   {(2-q)\sum_j p_j^{2-q}} \right).
\end{equation}
Note that if we replace $2-q$ with $q$ in Eq. \eqref{pdf1},
it becomes the pdf which Di Sisto {\it et al.} \cite{Sisto} 
and Bashkirov \cite{Bashkirov} have independently obtained
by modifying the treatment of the first version. 
We remark that the pdf of the original first version \cite{Tsallis88} is 
written in
\begin{equation}
   p_i = \frac{\qexp \left( -\beta^* E_i \right)}
          {\sum_i \qexp \left( -\beta^* E_i \right)},
\end{equation}
where $\beta^*$ is not the Lagrange multiplier associated with
the energy constraint.
From Eq. \eqref{pdf} and utilizing the identity \eqref{id-eq},
we readily obtain the relation among $\beta^*$, $\beta^{\rm (1)}$
and $\gamma^{\rm (1)}$ as
\begin{equation}
   \beta^* = \frac{\beta^{\rm (1)}}{1-(1-q) \gamma^{\rm (1)}}.
\end{equation}

%%%%%%%%%%%%%%%%%%%%%%%%%%%%%%%%%%%%%%%%%%%%%%%%%%%%%%%%%%%%%%%%%%%%%%%%%%%%
\section{Connections 
with the other versions of Tsallis' thermostatistics}
\vspace*{-7mm}

Let us now introduce the escort probability $P_i$ w.r.t. $p_i$ in 
the sense of Naudts' generalized 
thermostatistics \cite{Naudts@NEXT03,Naudts03}. 
For the $q$-exponential pdf, the $P_i$ can be 
written by
\begin{equation}
   P_i \equiv \frac{1}{Z_q} \cdot 
   \frac{d \qexp(x)}{dx} \Big\vert_{x= \qln (p_i)}
   = \frac{p_i^q}{Z_q} 
   \label{escort-pdf}
\end{equation}
where $Z_q$ is the normalization factor and we used
$d \qexp(x) / dx = \left[ \qexp(x) \right]^q$.
From the normalization of $P_i$, we have 
$Z_q = \sum_j p_j^q$. We see then that the escort probability of 
Eq. \eqref{escort-pdf} is nothing but the {\it so-called} $q$-escort 
probability
\cite{Tsallis98,escort}
\begin{equation}
    P_i = \frac{p_i^q}{\sum_j p_j^q},
\end{equation}
and that the average energy w.r.t. $P_i$ is the normalized $q$-average energy
\cite{Tsallis98}
\begin{equation}
    \ave{E_i}_{\rm P} \equiv  \sum_i  E_i P_i = U_q^{\rm (3)},
    \label{Uq}
\end{equation}
where $\ave{ \cdots }_{\rm P}$ stands for the average value 
w.r.t. the escort probability $P_i$.

Next let us confirm the original pdf of Eq. \eqref{pdf} is
also equivalent to that of the third version \cite{Tsallis98} in 
Tsallis' thermostatics. The key point is that $\gamma^{\rm (1)}$
is expressed not only in terms of $U^{\rm (1)}$ as Eq. \eqref{gamma}
but also in terms of $U_q^{\rm (3)}$.
By taking the average of the both sides of 
Eq. \eqref{scaled} w.r.t. $P_i$, we obtain
\begin{equation}
  \ave{ \qln \left( \frac{p_i}{\alpha}  \right) }_{\rm P}
  = -\beta^{\rm (1)} U_q^{\rm (3)} -\gamma^{\rm (1)},
  \label{aveP}
\end{equation}
and the l.h.s. is further expressed as
\begin{equation}
  \ave{ \qln \left( \frac{p_i}{\alpha}  \right) }_{\rm P}
  = 1-(2-q) S_q^{\rm N},
\end{equation}
where
\begin{equation}
S_q^{\rm N} \equiv \frac{S_q}{ \sum_j p_j^q}
 = \frac{1-\frac{\sum_i p_i}{ \sum_j p_j^q}}{1-q},
  \label{Sq^N}
\end{equation}
 is the normalized Tsallis entropy \cite{Landsberg,Rajagopal}.
We then have
\begin{equation}
  \gamma^{\rm (1)} = (2-q) S_q^{\rm N} -1-\beta^{\rm (1)} U_q^{\rm (3)}.
   \label{gamma1}
\end{equation}
Substituting this $\gamma^{\rm (1)}$ into Eq. \eqref{pdf} and
utilizing Eq. \eqref{id-eq}, we obtain
\begin{align}
  p_i &=
 \alpha \qexp \left( 1-(2-q)S_q^{\rm N} \right) \cdot
   \qexp \left( \frac{-\beta^{\rm (1)} (E_i-U_q^{\rm (3)})}
   {1+(1-q) \{ 1-(2-q)S_q^{\rm N} \} } 
    \; \right)
\nonumber \\
&= \frac{1}{\bar{Z}_q^{\rm (3)}}
   \; \qexp\left( \frac{-\beta^{\rm (1)} \sum_k p_k^q}{2-q}\; 
   (E_i - U_q^{\rm (3)}) \right),
  \label{pdf3}
\end{align}
where $\bar{Z}_q^{\rm (3)}$ is the $q$-generalized partition
function, and from the normalization of $p_i$, it can be written as
\begin{equation}
  \bar{Z}_q^{\rm (3)} 
   = \sum_i \qexp \left( \frac{-\beta^{\rm (1)} \sum_j p_j^{q}}{2-q} \;
     (E_i - U_q^{\rm (3)}) \right).
\end{equation}
From Eq. \eqref{pdf3} we readily confirm the following known 
relation \cite{Tsallis98}
\begin{equation}
  \bar{Z}_q^{\rm (3)} 
= \left[ \alpha 
  \qexp \left( 1-(2-q) S_q^{\rm N} \right) \right]^{-1}
= \left( \sum_i p_i^{q} \right)^{\frac{1}{1-q}}.
  \label{Z3}
\end{equation}
By comparing Eq. \eqref{Z1} with Eq. \eqref{Z3}, it follows that
\begin{equation}
  \bar{Z}_{q}^{\rm (1)}  = \bar{Z}_{q}^{\rm (3)}.
\end{equation}
This result is a consequence of ``$q \leftrightarrow 2-q$''-duality.

By the way, the pdf of the third version \cite{Tsallis98} is written by
\begin{align}
  p_i = \frac{1}{\bar{Z}_q^{\rm (3)} }
   \; \qexp \left( -\frac{\beta^{\rm (3)}}{\sum_j p_j^q}
\; (E_i - U_q^{\rm (3)}) \right),
 \label{p3}
\end{align}
which can be obtained as the solution of the following MaxEnt
problem
\begin{equation} 
 \frac{\delta}{\delta p_i} 
 \left(
    S_{q} - \beta^{\rm (3)} \frac{\sum_j p_j^q E_j}{\sum_k p_k^q} 
   -\gamma^{\rm (3)} \sum_j p_j
 \right) = 0,
  \label{MaxEnt3}
\end{equation}
where $\beta^{\rm (3)}$ and $\gamma^{\rm (3)}$ are the Lagrange multipliers
associated with the normalized $q$-average energy $U_q^{\rm (3)}$ and 
the normalization of the pdf, respectively.
From Eq. \eqref{MaxEnt3} it follows \cite{Plastino97,Yamano}
\begin{equation}
  \beta^{\rm (3)} = \frac{d S_q}{d U_q^{\rm (3)}}.
\end{equation}
By comparing Eq. \eqref{pdf3} with Eq. \eqref{p3} we find that 
the both pdfs are equivalent each other,
and that $\beta^{\rm (1)}$ and $\beta^{\rm (3)}$ are related by 
\begin{equation}
  \beta^{\rm (3)} = \frac{(\sum_j p_j^q)^2}{2-q} \beta^{\rm (1)}.
  \label{beta1-3}
\end{equation}

From Eq. \eqref{MaxEnt3} we have
\begin{equation} 
 \frac{q p_i^{q-1}-1}{1-q} 
 - \beta^{\rm (3)} \frac{ q p_i^{q-1}}{\sum_j p_j^q} 
        \left(E_i - U_q^{\rm (3)} \right)
   -\gamma^{\rm (3)} = 0.
  \label{MaxEnt3-rel}
\end{equation}
Multiplying the both sides of this equation by $p_i$
and taking summation, we obtain
\begin{equation}
  \gamma^{\rm (3)} = q S_q -1.
  \label{gamma3}
\end{equation}

We next consider the relations with the pdf for the normalized Tsallis 
entropy given in Eq. \eqref{Sq^N} \cite{Landsberg,Rajagopal}
\begin{align}
  p_i = \frac{1}{\bar{Z}_q^{\rm N(3)} }
   \; \qexp\left( -\beta^{\rm N(3)} \cdot \sum_j p_j^q \cdot
\; \left( E_i - U_q^{\rm (3)} \right) \right),
 \label{pN}
\end{align}
where
\begin{align}
  \bar{Z}_q^{\rm N(3)} =
   \left( \sum_j p_j^q \right)^{\frac{1}{1-q}},
\end{align}
and from the normalization of $p_i$, it follows that
\begin{align}
  \bar{Z}_q^{\rm N(3)} = 
   \sum_i \qexp \left( -\beta^{\rm N(3)} \cdot \sum_j p_j^q \cdot
\; \left( E_i - U_q^{\rm (3)} \right) \right).
\end{align}
$p_i$ can be obtained as the solution of the following MaxEnt
problem
\begin{equation} 
 \frac{\delta}{\delta p_i} 
 \left(
    S_{q}^{\rm N} - \beta^{\rm N(3)} \cdot 
         \frac{\sum_j p_j^q E_j}{\sum_k p_k^q} 
   - \gamma^{\rm N(3)} \cdot \sum_j p_j
 \right) = 0,
  \label{MaxEntN}
\end{equation}
where $\beta^{\rm N(3)}$ and $\gamma^{\rm N(3)}$ are the Lagrange multipliers
associated with the normalized $q$-average energy $U_q^{\rm (3)}$ and 
the normalization of the pdf, respectively.
From Eq. \eqref{MaxEntN} we have \cite{Plastino97,Yamano}
\begin{equation}
  \beta^{\rm N(3)} = \frac{d S_q^{\rm N}}{d U_q^{\rm (3)}},
\end{equation}
and by comparing Eq. \eqref{pdf3} with Eq. \eqref{pN} we find that 
the both pdfs are equivalent each other,
if 
\begin{equation}
  \beta^{\rm N(3)} = \frac{\beta^{\rm (1)}}{2-q} .
  \label{beta1-N}
\end{equation}

From Eq. \eqref{MaxEntN} it follows
\begin{equation} 
 \frac{1}{(\sum_j p_j^q)^2} \cdot 
  \left( \frac{q p_i^{q-1}-\sum_k p_k^q}{1-q} \right) 
 - \beta^{\rm N(3)} \cdot \frac{ q p_i^{q-1}}{\sum_j p_j^q} 
        \left(E_i - U_q^{\rm (3)} \right)
   -\gamma^{\rm N(3)} = 0.
  \label{MaxEntN-rel}
\end{equation}
Multiplying  the both sides of this equation by $p_i$
and taking summation, we obtain
\begin{equation}
  \gamma^{\rm N(3)} = (1-q) S_q^{\rm N} -1 = -\frac{1}{\sum_i p_i^q}.
  \label{gammaN}
\end{equation}

Until here, we have considered the pdfs
of the three different versions in which the combinations
of the entropies and average energies are: i) $S_{2-q}$ and $U^{\rm (1)}$;
ii) $S_q$ and $U_q^{\rm (3)}$; and iii) $S_q^{\rm N}$ and $U_q^{\rm (3)}$.
It is thus natural to consider the pdf for the combination of the normalized
Tsallis entropy with $q$ replaced by $2-q$,
\begin{equation}
   S_{2-q}^{\rm N} \equiv \frac{S_{2-q}}{\sum_j p_j^{2-q}},
\end{equation}
and $U^{\rm (1)}$.
The associated pdf can be written as
\begin{equation}
  p_i = \frac{1}{ \bar{Z}_{2-q}^{\rm N(1)}}\,
     \exp_{q} \left( -\frac{\beta^{\rm N(1)}}
     {2-q}\,\sum_i p_{i}^{2-q} \,\left( E_{i}-U^{\rm (1)} \right) \right)\ ,
\end{equation}
which can be obtained as the solution of the following MaxEnt problem
\begin{equation}
 \frac{\delta}{\delta \, p_i} \left( S_{2-q}^{\rm N} 
  - \beta^{\rm N(1)} \cdot \sum_j E_j \, p_j 
   -\gamma^{\rm N(1)} \cdot \sum_j p_j \right) =0 \ ,
\label{MaxEnt-N1}
\end{equation}
where $\beta^{\rm N(1)}$ and $\gamma^{\rm N(1)}$
are the Lagrange multipliers associated with the linear average energy
$U^{\rm (1)}$ and the normalization of the pdf, respectively. By
comparing this equation with Eq. \eqref{pdf1}, we find that the both pdf
are equivalent each other, if
\begin{equation}
  \beta^{\rm N(1)} = \frac{ \beta^{\rm (1)} }
              { \left( \sum_i p_i^{2-q} \right)^2 }\ ,
\end{equation}
and
\begin{equation}
\bar{Z}_{2-q}^{\rm (1)} = \bar{Z}_{2-q}^{\rm N(1)}.
\end{equation}
From Eq. \eqref{MaxEnt-N1} it follows
\begin{equation}
- \frac{1}{(q-1) \sum_j p_j^{2-q} }\,\left( 1-(2-q)\,
  \frac{\sum_k p_k}{ \sum_{\ell} p_{\ell}^{2-q}} \, p_i^{1-q} \right)
   -\beta^{\rm N(1)} \, E_i -\gamma^{\rm N(1)} =0 \ .
\end{equation}
Multiplying both sides of this equation by $p_{_i}$ and taking the
summation, we obtain
\begin{equation}
 \gamma^{\rm N(1)} = (q-1)\,S_{2-q}^{\rm N} -1
    = -\frac{1}{\sum_i p_{i}^{2-q}}.
\end{equation}

Summing up, the pdfs of the different versions of Tsallis' thermostatistics
are related one another.
Note that $U_q^{\rm (3)}$ is automatically introduced
as the average energy w.r.t. the escort probability. This
is a consequence of the ``$q \leftrightarrow 1/q$''-duality. 
In other words, $U_q^{\rm (3)}$ is accompanied with $U^{\rm (1)}$.
From Eqs. \eqref{gamma} and \eqref{gamma1}, we see that they are related by
\begin{equation}
  S_{2-q} -\left( \frac{ \beta^{\rm (1)} }{2-q} \right) U^{\rm (1)}
   = S_q^{\rm N} -\beta^{\rm N(3)} U_q^{\rm (3)}.
   \label{U1-U3}
\end{equation}
By taking the derivative of both sides of this equation w.r.t. 
$\beta^{^{(1)}}$, we obtain
\begin{equation}
  (1-q)\,\beta^{\rm (1)} \,\frac{d\,U^{\rm (1)}}{d\,\beta^{\rm (1)}}
 = U^{\rm (1)}-U_q^{\rm (3)}\ .
 \label{dU1-U3}
\end{equation}
In our opinion,
it is thus meaningless asking which of the two average energies is correct. 
They cannot exclude each other.

%%%%%%%%%%%%%%%%%%%%%%%%%%%%%%%%%%%%%%%%%%%%%%%%%%%%%%%%%%%%%%%%%%%%%%%%%%
\section{Generalized Massieu potential and associated
Legendre structures}
\vspace*{-7mm}

Let us first remind you of some basic relations concerning with
Massieu's potential \cite{Callen} in the standard BG 
thermostatistics.
Massieu's potential $\Phi$ is defined as the Legendre transform of 
the standard BG entropy $S(U)$ which is a function of 
internal energy $U$,
\begin{align}
  \Phi(\beta) \equiv S - \beta U.
  \label{Massieu}
\end{align}
Massieu's potential is thus a function of the Lagrange multiplier $\beta$
associated with energy constraint,
whereas Helmholtz free energy $F$ is defined as
the Legendre transform of the internal energy $U(S)$ which is a function
of $S$,
\begin{align}
  F(T) \equiv U - T S,
  \label{F}
\end{align}
and consequently $F$ is a function of temperature $T$.
Since the temperature $T$ and the Lagrange multiplier $\beta$
are related by $T= 1/\beta$ in the standard BG thermostatistics,
Massieu's potential is related with free energy as
\begin{equation}
  \Phi = -\frac{F}{T}.
\end{equation}

By differentiating the both side of Eq. \eqref{Massieu} and
utilizing the relation
\begin{align}
  \frac{d S(\beta)}{d \beta} 
        = \beta \frac{d U(\beta)}{d \beta},
  \label{dS-dU}
\end{align}
we readily obtain
\begin{align}
  \frac{d \Phi}{d \beta} = -U.
  \label{dMp}
\end{align}
Eqs. \eqref{Massieu}, \eqref{dS-dU}, and \eqref{dMp} are basic 
 relations concerning with Massieu's potential
in the standard BG thermostatistics. 

We next review the relation between Massieu's potential
and the Lagrange multiplier $\gamma$ associated with
the normalization of probability in the standard MaxEnt procedure,
\begin{equation} 
 \frac{\delta}{\delta p_i} 
 \left(
    S - \beta \sum_j p_j E_j - \gamma \sum_j p_j
 \right) = 0.
  \label{BG-MaxEnt}
\end{equation}
Its solution is the well-known BG pdf
\begin{equation}
  p_i^{\rm BG} = \exp \left( -\beta E_i - \gamma-1 \right)
   = \frac{1}{Z} \exp \left( -\beta E_i \right),
 \label{BG-pdf}
\end{equation}
where the partition function $Z = \exp \left( 1+\gamma \right)$ is
introduced.
By substituting Eq. \eqref{BG-pdf} into the BG entropy $S$,
we obtain
\begin{equation}
  S = -\sum_i p_i^{\rm BG} \ln p_i^{\rm BG}
   =   \sum_i p_i^{\rm BG} \left( \beta E_i + \gamma+1 \right)
   = \beta U + \gamma + 1.
\end{equation}
Comparing this with Eq. \eqref{Massieu}, we have the known
relations
\begin{equation}
  \Phi = 1+\gamma = \ln Z.
  \label{Phi}
\end{equation}
We thus see that the Lagrange multiplier $\gamma$ is related to
Massieu' potential $\Phi$, and that Eq. \eqref{dMp} is
equivalent to the well-known relation $d\ln Z/d \beta = -U$.

Let us now focus on the generalizations
of these basic relations within Tsallis' thermostatistics.
Naudts \cite{Naudts@NEXT03} has already shown that there exists
a generalized (Helmholtz) free energy
associated with the average energy w.r.t. the escort probability 
in his generalized thermostatistics, which contains Tsallis' 
thermostatistics as a special case.
For a generalized entropy based on a deformed logarithmic function,
the Lagrange multiplier associated with an energy constraint
is generally not equivalent to the inverse temperature $1/T$.
In our formalism it is thus appropriate to introduce a generalized
Massieu potential, which is a function of 
the Lagrange multiplier $\beta^{\rm (1)}$ or $\beta^{\rm N(3)}$, 
instead of the generalized free energy, which is a function of temperature.
Following the same method used by Naudts to derive the generalized 
free energy \cite{Naudts@NEXT03}, let us derive the
generalized Massieu potential. 

By utilizing Eqs. \eqref{alpha} and \eqref{id-eq}, 
we rewrite the original pdf of Eq. \eqref{pdf} as
\begin{align}
  p_i &= \qexp \big( \qln (\alpha) \big) \cdot
      \qexp \left( -\beta^{\rm (1)} E_i-\gamma^{\rm (1)} \right) \nonumber \\
    &= \qexp \left( \frac{-1}{2-q} 
        \left( \beta^{\rm (1)} E_i+1+\gamma^{\rm (1)} \right)  \right)
    = \qexp \left( -\beta^{\rm N(3)} E_i - \Phi_q^{\rm N(3)}  \right),
   \label{pi}
\end{align}
where in the last step we introduced the quantity
\begin{equation}
      \Phi_q^{\rm N(3)} \equiv \frac{1 +\gamma^{\rm (1)}}{2-q},
   \label{Phiq}
\end{equation}
which reduces to Eq. \eqref{Phi} in the limit of $q \to 1$.

By differentiating the both sides of $\sum_i p_i = 1$ 
 w.r.t. $\beta^{\rm N(3)}$,
and utilizing Eqs. \eqref{escort-pdf}, \eqref{Uq} and \eqref{pi}, we have
\begin{align}
 0 &= \sum_i \frac{d p_i}{d \beta^{\rm N(3)}}
   = -\sum_i \left( E_i+\frac{d \Phi_q^{\rm N(3)}}{d \beta^{\rm N(3)}} \right)
       \frac{d \qexp (x)}{d x} 
\Big\vert_{ x = -\beta^{\rm N(3)} E_i-\Phi_q^{\rm N(3)} }
   \nonumber\\
   &= - \sum_i 
  \left( E_i + \frac{d \Phi_q^{\rm N(3)} }{d \beta^{\rm N(3)}} \right) Z_q P_i
   = -Z_q \left(
   U_q^{\rm (3)} + \frac{d \Phi_q^{\rm N(3)}}{d \beta^{\rm N(3)}} \right).
\end{align}
We then obtain
\begin{equation}
   \frac{d \Phi_q^{\rm N(3)} }{d \beta^{\rm N(3)}}  = -U_q^{\rm (3)}.
   \label{Mp-derive}
\end{equation}
By comparing this relation with Eq. \eqref{dMp},
we find that $\Phi_q^{\rm N(3)}$ is the generalized Massieu potential
associated with the escort average energy $U_q^{\rm (3)}$.
In the limit of $q \to 1$, Eq. \eqref{Mp-derive} of course reduces to
Eq. \eqref{dMp}.

Now a natural question arises at this point: what are the generalizations of 
Eqs. \eqref{Massieu} and \eqref{dS-dU}? In other words, what is a generalized
entropy whose Legendre transform is $\Phi_q^{\rm N(3)}$?
From Eq. \eqref{gamma1} and the definition of $\Phi_q^{\rm N(3)}$ given by 
Eq. \eqref{Phiq}, we obtain
\begin{equation}
\Phi_q^{\rm N(3)} =  S_q^{\rm N} - \beta^{\rm N(3)} \cdot U_q^{\rm (3)},
   \label{Phi_N}
\end{equation}
which shows that the generalized Massieu potential
$\Phi_q^{\rm N(3)} \left( \beta^{\rm N(3)} \right)$ is 
the Legendre transform 
of $S_q^{\rm N} \left( U_q^{\rm (3)} \right)$.
By differentiating the both sides of Eq. \eqref{Phi_N} w.r.t. 
$\beta^{\rm N(3)}$ and utilizing Eq. \eqref{Mp-derive}, we obtain
\begin{equation}
  \frac{d S_q^{\rm N} }{d \beta^{\rm N(3)}}
  = \beta^{\rm N} \; \frac{d U_q^{\rm (3)}}{d \beta^{\rm N(3)}}.
 \label{derive}
\end{equation}
Eqs. \eqref{Phi_N} and \eqref{derive} 
lead us to consider  $S_q^{\rm N}$
as the generalized entropy associated with 
the Massieu potential $\Phi_q^{\rm N(3)}$.

Now we'd like to point out some observations.
Firstly, it is worth noting that $\Phi_q^{\rm N(3)}$ is not 
associated with $U^{\rm (1)}$ but associated with $U_q^{\rm (3)}$.
In fact, from Eq. \eqref{gamma}, $\Phi_q^{\rm N(3)}$ can be also expressed as
\begin{equation}
\Phi_q^{\rm N(3)} =  S_{2-q} - \frac{ \beta^{\rm (1)}}{2-q} \cdot U^{\rm (1)}.
   \label{Phiq-U1}
\end{equation}
By taking the derivative of the both sides of this equation
w.r.t. $\beta^{\rm (1)}$
and utilizing Eq. \eqref{S2-q-U1}, we have
\begin{equation}
   \frac{d \Phi_q^{\rm N(3)} }{d \beta^{\rm (1)} }  
  = -\frac{U^{\rm (1)}}{2-q} 
   + \left( \frac{1-q}{2-q} \right) 
      \beta^{\rm (1)} \frac{d U^{\rm (1)}}{d \beta^{\rm (1) }}.
   \label{Mp-derive1}
\end{equation}
This is not a form invariant generalization of Eq. \eqref{dMp},
whereas Eq. \eqref{Mp-derive} is a natural generalization.

However, we can construct an appropriate generalization
in order to overcome this difficultly
by utilizing the fact that  $U^{\rm (1)}$ and $ U_q^{\rm (3)}$
are related to each other as shown in Eq. \eqref{U1-U3}
or Eq. \eqref{dU1-U3}.
Let us define the Massieu potential associated
with $U^{\rm (1)}$ as
\begin{equation}
   \Phi_{2-q} \equiv \Phi_q^{\rm N(3)} 
   - \left( \frac{1-q}{2-q} \right) \beta^{\rm (1)} \cdot U^{\rm (1)}
   = \frac{1 +\gamma^{\rm (1)}}{2-q}
      - \left( \frac{1-q}{2-q} \right) \beta^{\rm (1)} \cdot U^{\rm (1)}.
   \label{Phi2-q}
\end{equation}
Substituting Eq. \eqref{gamma} or Eq. \eqref{Phiq-U1} into this equation 
leads to
\begin{equation}
        \Phi_{2-q} = S_{2-q}-\beta^{\rm (1)} \cdot U^{\rm (1)},
\end{equation}
which shows that the generalized Massieu potential
$\Phi_{2-q} \left( \beta^{(1)} \right)$ is 
the Legendre transform 
of $S_{2-q} \left( U^{\rm (1)} \right)$.
Furthermore from Eqs. \eqref{Mp-derive1} and \eqref{Phi2-q},
we obtain
\begin{equation}
   \frac{d \Phi_{2-q} }{d \beta^{\rm (1)} }  
  = -U^{\rm (1)},
   \label{Mp-derive2}
\end{equation}
which can be also derived by utilizing Eq. \eqref{dU1-U3}
in the same way as we have derived Eq. \eqref{Mp-derive}.

Secondly, in the literature, the generalized free 
energy $F_q^{\rm (3)}$ \cite{Tsallis98} 
associated
with the normalized $q$-average energy $ U_q^{\rm (3)}$ is known as
\begin{equation}
F_q^{\rm (3)} \equiv  U_q^{\rm (3)} 
          - \frac{1}{\beta^{\rm (3)}} \cdot S_q^{\rm (3)}.
   \label{Fq3}
\end{equation}
The corresponding Massieu potential can be written as
\begin{equation}
\Phi_q^{\rm (3)} \equiv  -\beta^{\rm (3)} F_q^{\rm (3)} =
    S_q^{\rm (3)}  - \beta^{\rm (3)} \cdot U_q^{\rm (3)}.
   \label{Phiq3}
\end{equation}
For the version which utilizes the standard linear energy
average $U^{\rm (1)}$ as the constraint in the MaxEnt procedure, 
the Lagrange multiplier $\gamma^{\rm (1)}$, which is associated with
the normalization of the pdf, is related with the generalized Massieu potential
in the same way of the standard BG thermostatistics.\\
Note that $\Phi_q^{\rm N(3)}$ has no relation to $\gamma^{\rm N(3)}$.
In fact both the generalized Massieu potentials $\Phi_q^{\rm N(3)}$
and $\Phi_{2-q}$ are related only with $\gamma^{\rm (1)}$.
However, the situation is different for the versions utilizing
the normalized $q$-average energy $U_q^{\rm (3)}$, for which
the reference energy is forced to shift from zero to $U_q^{\rm (3)}$
\cite{Tsallis98}. 
Consequently the Lagrange multipliers $\gamma^{\rm (3)}$ and 
$\gamma^{\rm N(3)}$ are related only with $S_q$ 
(as shown in Eq. \eqref{gamma3}) and 
$S_q^{\rm N}$ (as Eq. \eqref{gammaN}), respectively.\\
It is thus difficult to obtain
the $\Phi_q^{\rm (3)}$ or $F_q^{\rm (3)}$ by applying our method,
which is based on the relation between the Lagrange multiplier
associated with the normalization of pdf and Massieu' potential.

\section{Conclusions}
\vspace*{-7mm}

Based on the Tsallis' thermostatistics with
the standard linear average energy $U^{\rm (1)}$ and $S_{2-q}$,
we have explored the connections among the pdfs of the two different
kinds of Tsallis' formalism: one employs $U^{\rm (1)}$,
and the other employs the normalized $q$-average energy $U_q^{\rm (3)}$ 
as the energy constraint.\\
We have shown the relations among the Lagrange multipliers associated
with energy constraints of the different versions.
It is revealed that the standard linear average energy $U^{\rm (1)}$
and the normalized $q$-average energy $U_q^{\rm (3)}$ are related 
to each other.\\
Furthermore we have studied the relevant thermodynamic Legendre relations
concerning with the Lagrange multiplier $\gamma^{\rm (1)}$ associated
with the normalization of the pdf.
By utilizing the relation between $U^{\rm (1)}$ and $U_q^{\rm (3)}$, 
we have constructed the generalized Massieu potential 
associated with either $U^{\rm (1)}$ or $U_q^{\rm (3)}$.

\section*{Acknowledgments}
\vspace*{-7mm}
This work was started when one of the author (T.W.) stayed
in Torino, Italy.
He wishes to thank the department of Physics, 
Politecnico di Torino for their hospitality, and
the Japanese Ministry of Education, 
Culture, Sports, Science and Technology for the Overseas Research Fellowship.


\begin{thebibliography}{}
\vspace*{-7mm}
%
\bibitem{Tsallis88}
C. Tsallis,
Possible generalization of Boltzmann-Gibbs statistics,
J. Stat. Phys. \textbf{52} (1988) 479-487.

\bibitem{Curado91}
E.M.F. Curado, and C. Tsallis,
Generalized statistical mechanics: connection with thermodynamics,
J. Phys. A: Math. Gen. \textbf{24} (1991) L69-72.

\bibitem{Tsallis98}
C. Tsallis, R.S. Mendes, A.R. Plastino,
The role of constraints within generalized nonextensive statistics,
Physica A \textbf{261} (1998) 534-554.

\bibitem{Callen}
H.B. Callen, \textit{Thermodynamics and an Introduction to Thermostatistics},
2nd ed. (Wiley New York 1985)

\bibitem{NEXT01}
G. Kaniadakis, M. Lissia, A. Rapisarda (editors),
\textit{Proc. of the international school and workshop on non extensive
thermodynamics and physical applications (NEXT2001)}
Physica A \textbf{305} Issues 1-2 (2001).

\bibitem{NEXT03}
G. Kaniadakis, M. Lissia (editors),
\textit{Proc. of the 2nd Sardinian International Conference on News and
Expectations in Thermostatistics (NEXT2003)}
Physica A \textbf{340} Issues 1-3 (2004).

\bibitem{Refs}
The compact background and review by Tsallis himself
can be obtained from
arXiv: cond-mat/0312699. \\
The updated bibliography can be obtained from the following URL: 
http://tsallis.cat.cbpf.br/biblio.htm


\bibitem{escort}
C. Beck and F. Schl\"{o}gl, {\it Thermodynamics of Chaotic Systems}
(Cambridge U. P. Cambridge, 1993) p 53.

%\bibitem{Beck-Cohen}
%C. Beck and E.G.D. Cohen, 
%Superstatistics, Physica A \textbf{322} (2003) 267-275.
%
%\bibitem{Beck}
%C. Beck,
%Superstatistics, escort distributions, and applications,
%arXiv: cond-mat/0312134.

\bibitem{Naudts-CSF}
J. Naudts, Dual description of nonextensive ensembles, 
Chaos, Solitons \& Fractals \textbf{13} (2002) 445-450.

\bibitem{Raggio}
G.A. Raggio,
Equivalence of two thermostatistical formalisms based on 
the Havrda \& Charvat-Daroczy-Tsallis entropies,
arXiv: cond-mat/9908207.

\bibitem{0th-law}
S. Mart\'inez, F. Pennini, A. Plastino,
Thermodynamics' 0-th-Law in a nonextensive scenario,
Physica A \textbf{295} (2001) 416-424.

\bibitem{Abe-Martinez01}
S. Abe, S. Mart\'inez, F. Pennini, A. Plastino,
Nonextensive thermodynamic relations,
Phys. Lett. A \textbf{281} (2001) 126-130.

\bibitem{Martinez00}
S. Mart\'inez, F. Nicol\'as, F. Pennini, and A. Plastino,
Tsallis' entropy maximization procedure revisited,
Physica A \textbf{286} (2000) 489-502.

\bibitem{Abe01}
S. Abe, S. Mart\'nez, F. Pennini, and A. Plastino,
Classical gas in nonextensive optimal Lagrange multipliers formalism,
Phys. Lett. A \textbf{278} (2001) 249-254.

\bibitem{Casas02}
M. Casas, S. Mart\'inez, F. Pennini, and A. Plastino,
Thermodynamics and the Tsallis variational problem,
Physica A \textbf{305} (2002) 41-47.

\bibitem{Landsberg}
P. Landsberg, and V. Vedral,
Distributions and channel capacities in generalized statistical mechanics,
Phys. Lett. A \textbf{247} (1998) 211-217.

\bibitem{Rajagopal}
A.K. Rajagopal, and S. Abe,
Implications of Form Invariance to the Structure of Nonextensive Entropies,
Phys. Rev. Lett. \textbf{83}  (1999) 1711-1714.

\bibitem{escort-Sq}
C. Tsallis, and E. Brigattai,
Nonextensive statistical mechanics: A brief introduction,
Continuum Mech. Thermodyn. \textbf{16} (2004) 223-235.

\bibitem{Abe04}
S. Abe, and G.B. Bagci,
Constraints and relative entropies in nonextensive statistical mechanics,
arXiv:~cond-mat/0404253.

\bibitem{Naudts02}
J. Naudts, Deformed exponentials and logarithms in generalized 
thermo-statistics, Physica A \textbf{316} (2002) 323-334.

\bibitem{Sisto}
R.P. Di Sisto, S. Mart\'inez, R.B. Orellana, A.R. Plastino, A. Plastino,
General thermostatistical formalisms, invariance under uniform spectrum
translations, and Tsallis $q$-additivity,
Physica A \textbf{265} (1999) 590-613.

\bibitem{Bashkirov}
A.G. Bashkirov,
On maximum entropy principle, superstatistics, power-law distribution
and Renyi parameter,
Physica A \textbf{340} (2004) 153-162.

\bibitem{Baldovin}
F.~Baldovin, and A.~Robledo,
Nonextensive Pesin identity, Exact re-normalization
group analytical results for the dynamics at the edge of 
chaos of the logistic map,	
arXiv:~cond-mat/0304410 the paragraph around Eq. (8).

\bibitem{Naudts@NEXT03}
J. Naudts,
Generalized thermostatistics based on deformed exponential and 
logarithmic functions, Physica A \textbf{340} (2004) 32-40.
arXiv:~cond-mat/0311438.

\bibitem{Naudts03}
J. Naudts,
Generalized thermostatistics and mean field theory,\\
Physica A \textbf{332} (2003) 279-300, arXiv:~cond-mat/0211444.  


\bibitem{Abe}
S. Abe,
Generalized entropy optimized by a given arbitrary distribution,\\
J. Phys. A: Math. Gen. \textbf{36} (2003) 8733-8738.

\bibitem{Kaniadakis}
G. Kaniadakis, M. Lissia, and A.M. Scarfone,
Deformed logarithms and entropies,
Physica A \textbf{340} (2004) 41-49.

\bibitem{Kaniadakis04}
G. Kaniadakis, M. Lissia, and A.M. Scarfone,
Two-parameter deformations of logarithm, exponential,
and entropy: a consistent framework for generalized statistical
mechanics, arXiv: cond-mat/0409683.

\bibitem{Plastino97}
A. Plastino, and A.R. Plastino,
On the universality of thermodynamics' Legendre transform structure,	
Phys. Lett. A \textbf{226} (1997) 257-263.

\bibitem{Yamano}
T. Yamano,
On the robust thermodynamical structures against arbitrary entropy form 
and energy mean value,
Eur. Phys. J. \textbf{B 18} (2000) 103-106.



\end{thebibliography}
\end{document}